\documentclass[aps,pre,a4paper,10pt,twocolumn,longbibliography,superscriptaddress,nofootinbib,bibnotes,notitlepage]{revtex4-2}
\usepackage{color}
\definecolor{red}{rgb}{0.75,0,0}
\definecolor{blue}{rgb}{0,0,0.75}
\definecolor{green}{rgb}{0,0.5,0}
\usepackage{amsmath}
\usepackage{amssymb}
\usepackage{bm}
\usepackage{xcolor}
\usepackage{graphicx}
\usepackage{datetime}
\longdate
\usepackage{comment}

\usepackage[normalem]{ulem}

\makeatletter
\def\maketitle{
	\@author@finish
	\title@column\titleblock@produce
	\suppressfloats[t]}
\makeatother

\def\be{\begin{equation}}
\def\ee{\end{equation}}
\def\bea{\begin{eqnarray}}
\def\eea{\end{eqnarray}}

\def\besub{\begin{subequations}}
\def\eesub{\end{subequations}}

\def\bwd{\begin{widetext}}
\def\ewd{\end{widetext}}

\newcommand{\qq}{\begin{eqnarray}}
\newcommand{\qqq}{\end{eqnarray}}

\begin{document}
\title{Screw Symmetry, Chiral Hydrodynamics and Odd Instability in Active Cholesterics}

\author{Gareth P. Alexander}
\email{G.P.Alexander@warwick.ac.uk}
\affiliation{Department of Physics, Gibbet Hill Road, University of Warwick, Coventry CV4 7AL, United Kingdom}

\author{S. J. Kole}
\email{sjkole@uchicago.edu}
\affiliation{Pritzker School of Molecular Engineering, University of Chicago, Chicago, Illinois 60637, USA}
\affiliation{DAMTP, Centre for Mathematical Sciences, University of Cambridge, Wilberforce Rd, CB3 0WA Cambridge, United Kingdom}

\author{Ananyo Maitra}
\email{nyomaitra07@gmail.com}
\affiliation{{Laboratoire de Physique Th\'eorique et Mod\'elisation, CNRS UMR 8089, CY Cergy Paris Universit\'e, F-95032 Cergy-Pontoise Cedex, France}}
\affiliation{Sorbonne Universit\'{e} and CNRS, Laboratoire Jean Perrin, F-75005, Paris, France}

\author{Sriram Ramaswamy}
\email{sriram@iisc.ac.in}
\affiliation{Centre for Condensed Matter Theory, Department of Physics, Indian Institute of Science, Bangalore 560 012, India}
\affiliation{International Centre for Theoretical Sciences, Tata Institute of Fundamental Research, Bangalore 560 089, India}

\date{\today}

\def\corrAuthor{Gareth P. Alexander}
\def\corrEmail{g.p.alexander@warwick.ac.uk}

\begin{abstract}
Active cholesterics are chiral in both their structure, which has continuous screw symmetry, and their active stresses, which include contributions from torque dipoles. Both expressions of chirality give rise to curl forces in the hydrodynamics, which we derive from the active Ericksen-Leslie equations using a geometric approach. This clarifies the hydrodynamics of continuous screw symmetry and provides an example of generalised odd elastic forces that originate from an equilibrium free energy. We discuss also the nonlinear structure of the active hydrodynamics in terms of the Eulerian displacement field of the cholesteric pseudolayers. For the active instability, screw symmetry generates a contribution of chiral activity to the linearised pseudolayer hydrodynamics that is absent in materials with chiral activity but achiral structure. When the two forms are sufficiently antagonistic, this term produces a new active instability with threshold and characteristic wavevector distinct from those of the active Helfrich-Hurault instability in chiral active smectics. Finally, we comment on the isotropic chiral hydrodynamics of materials with three-dimensional screw symmetry. 
\end{abstract}

\maketitle

\section{Introduction}
\label{sec:intro}

Active matter theories describe the physics of non-equilibrium systems in which the scale of the drive is smaller than the minimal lengthscale at which the theoretical description is valid and the driving mechanism is not spatially correlated. That is, unlike traditional non-equilibrium systems in which the driving mechanism explicitly breaks rotation symmetry, here it only breaks time-reversal symmetry. The physics of spontaneous breaking of various spatial symmetries in these systems has therefore been a fruitful area of research over the last three decades~\cite{TonTuRam,SRrev,RMP,bowick2022,maitra2025}. 

Chiral forms of active driving have taken prominence in recent years; these include spinning particles, helical swimming, microscopic torque dipoles and non-reciprocally actuated robotic assemblages. They show a range of macroscopic phenomena including odd viscous and elastic moduli~\cite{fruchart2023,Mietke,kole2021,kole2024}, unidirectional edge flows~\cite{li2024}, self-organised locomotion~\cite{ishimoto2022,ishimoto2023}, stirring~\cite{kole2025}, and adaptive locomotion~\cite{veenstra2025}. The majority of this has considered materials that are chiral only through their activity so that the interplay between chiral activity and chiral structure remains relatively unexplored. Chiral forms of passive materials are extremely rich, displaying a huge range of metastable states, geometric frustration and proliferation of topological defects~\cite{renn1988,wright1989,matsumoto2009,machon2017,pollard2019,wu2022}. Cholesteric liquid crystals provide a basic setting for the interplay between chiral activity and chiral structure, where active stresses of torque dipole form influence the helical director field. 

In terms of their spontaneous symmetry breaking, liquid crystals occupy a middle ground in that they remain invariant under some continuous spatial transformations and not others, thus leading to spectacular symmetry-mandated mechanics and dynamics. In both cholesterics and smectics the ground states are periodic along one spatial direction and the large scale mechanics of both is therefore a compressional elasticity along that direction and a bending elasticity perpendicular to it~\cite{lubensky1972,deGennesProst}. A hallmark of this is that both cholesterics and smectics exhibit the same mechanical Helfrich-Hurault instability~\cite{TranRMP}. At the same time, the spontaneous symmetry breaking of the two phases is distinct: cholesteric states are formed by aligning chiral molecules that, due to microscopic geometric frustration, organise in a helical structure; smectics have a one-dimensional periodic density modulation. While both structures are periodic, an arbitrary translation of the former can be compensated by a rotation; there is no such invariance in the latter. This distinction leads to differences in flow phenomena~\cite{helfrich1969} at subleading order in wavenumbers~\cite{lubensky1972}. 

In an earlier article \cite{kole2021}, we showed that activity allows chirality to play an important role in the hydrodynamics \cite{maitra2025}, leading to an anti-ferromagnetically organised columnar vortex lattice array in response to a periodic undulation of the layer (or pseudolayer) structure. However, we argued that the structural difference between cholesteric and smectic phases does not affect the small-wavenumber, low-frequency physics, i.e., active smectic A* and cholesteric phases are hydrodynamically equivalent. In this article, we reexamine the hydrodynamics of active cholesteric phases using a geometric approach to the Ericksen-Leslie equations that we adapt from that used for the free energy by Radzihovsky \& Lubensky~\cite{radzihovsky2011}. The screw symmetry of cholesterics is manifested in a coupling of fluid vorticity to the pseudolayer dynamics and a corresponding antisymmetric contribution to the hydrodynamic stress~\cite{lubensky1972}. We review the origin of this and show it produces a curl force of generalised odd elastic form but, in contrast to two-dimensional odd elasticity~\cite{fruchart2023}, here it originates from an equilibrium free energy. Crucially, the vorticity coupling term allows a chiral active force density to participate in the linear instability of the pseudolayered structure in the Stokesian regime, albeit at subleading order in wavenumbers. This changes the character of the instability with the threshold value of contractile activity for an active Helfrich-Hurault~\cite{kole2021,whitfield2017} in systems confined along the pitch direction and the most unstable transverse wavevector, both depending explicitly on the chiral active force density. Further, for sufficiently strong chiral activity with opposite handedness to the cholesteric helix it produces a new active instability with characteristics distinct from those of Helfrich-Hurault and ultimately leading to a breakdown of the cholesteric structure at the level of the pitch. 

The remainder of this paper is organised as follows. In Section~\ref{sec:screw_symmetry} we describe in less technical terms the chiral mechanics of active cholesterics and the nature of their generalised odd elasticity. In Section~\ref{sec:cholesteric} we obtain the hydrodynamics of active cholesterics through a geometric analysis of the Ericksen-Leslie equations adapting the methodology of Ref.~\cite{radzihovsky2011}. In Section~\ref{sec:nonlinear} we present the nonlinear hydrodynamic theory of active cholesterics in terms of the Eulerian displacement field. In Section~\ref{sec:linear_cholesteric} we analyse the linearised theory of cholesterics and identify their active instabilities, with focus on the rotation-translation coupling effects of screw symmetry. In Section~\ref{sec:3d_chiral_fluid} we examine the linear isotropic hydrodynamics of a material with three-dimensional rotation-translation couplings. Finally, in Section~\ref{sec:discussion} we end with a discussion.

\section{Chiral Mechanics and Screw Symmetry in Cholesterics}
\label{sec:screw_symmetry}

Chiral mechanical forces have been identified in a range of continuum materials and mesophases~\cite{fruchart2023,maitra2025}. A basic example is the odd elasticity of two-dimensional isotropic solids, which support a chiral force density $K_o \,{\bf J} \cdot \nabla^2 {\bf u}$, where ${\bf J}$ is the generator of (counterclockwise) rotations and ${\bf u}$ is the Eulerian displacement field. In three-dimensional active layered mesophases~\cite{kole2021}, the chiral mechanics are a curl force $\zeta_c \nabla \times (\nabla^2 u \,{\bf N})$, where here ${\bf N}$ is the layer normal and $u$ is again the Eulerian displacement field (along the layer normal). The same force is also present in active chiral membranes~\cite{al-izzi2023} and in both cases it may be viewed as originating from the active stress of microscopic torque-dipoles~\cite{furthauer2012,furthauer2013}. The chiral mechanics of active polar columnar phases contains forces analogous to both the curl force of active layered mesophases and the odd elasticity of two-dimensional solids~\cite{kole2024}. In this case, the three-dimensionality of the columnar phase, together with the viscous part of its mechanics, leads to an odd dynamics with different characteristics from those that arise for truly two-dimensional materials. 

\begin{figure*}[t]
    \centering
    \includegraphics[width=0.98\textwidth]{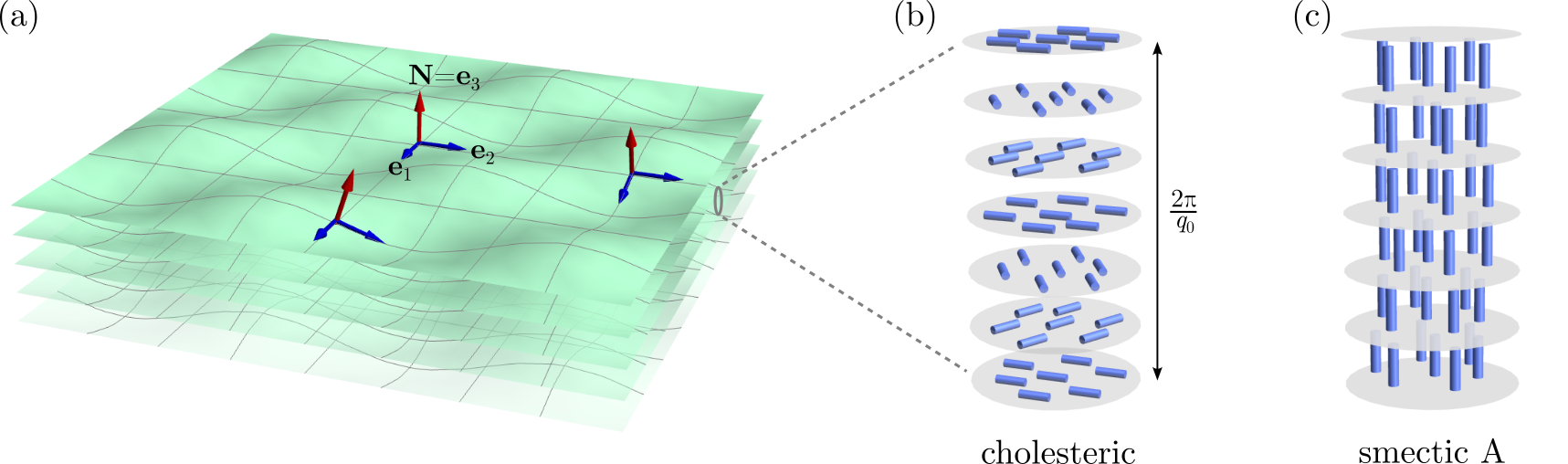}
    \caption{(a) Local structure of a (pseudo)layered material with adapted orthonormal frame. The (pseudo)layers correspond to a (local) discrete translational symmetry along the pitch axis or layer normal ${\bf N}$. (b) In a cholesteric each pseudolayer corresponds to a full $2\pi$ rotation of the director field, shown by the blue cylinders. The grey discs provide a guide to the eye. The structure has continuous screw symmetry (here right-handed) along the pitch axis. (c) In smectic A the director points along the layer normal and the layers correspond to a genuine density modulation. There is continuous rotational symmetry about the layer normal but only discrete translational symmetry along it.} 
    \label{fig:smectic_cholesteric}
\end{figure*}

Although the chiral mechanics of odd elasticity and active layered mesophases have their own distinctiveness, they can be viewed as more alike mathematically by recasting (part of) odd elasticity as a two-dimensional curl force. Taking a Helmholtz decomposition of the displacement field, ${\bf u} = \nabla \phi + {\bf J} \cdot \nabla A + {\bf u}_0$, where $\nabla \phi$ is irrotational, ${\bf J} \cdot \nabla A$ is incompressible and ${\bf u}_0$ denotes the harmonic component, the force density is $K_o \,{\bf J} \cdot \nabla (\nabla^2\phi) - K_o \nabla (\nabla^2 A)$ and the irrotational part is explicitly a curl, while the incompressible part is a gradient. The harmonic component does not enter the force density but does contribute to torques when it represents a non-trivial element of cohomology, see e.g.~\cite{lapa2014,ganeshan2017}. 

Cholesterics have a one-dimensional periodic modulation that can be described as a pseudolayer structure. In active cholesterics, this allows for the same curl force $\zeta_c \nabla \times (\nabla^2 u \,{\bf N})$ as arises in general chiral active layered mesophases. However, the symmetry of cholesterics is not the same as that of an explicitly layered mesophase such as a smectic. In cholesterics the symmetry is that of a screw axis that couples rotations about the helical axis with translations along it; in smectic A, rotations about the layer normal with no translation are symmetries; see Fig.~\ref{fig:smectic_cholesteric}. To say more, the ground states of both cholesterics and smectic A are described in terms of a linear phase function $\phi = z$, or any equivalent to it under a Euclidean motion. However, this phase function alone does not fully capture their symmetries. The smectic A ground state is a density modulation, $\rho = \rho_0 + \delta \rho \cos q_{\textrm{sm}}z$, that has full rotational symmetry about the $z$-axis (layer normal) and discrete translational symmetry (by the layer spacing $2\pi/q_{\textrm{sm}}$) along it. The cholesteric ground state is the director field ${\bf n} = \cos q_0 z \,{\bf e}_x + \sin q_0 z \,{\bf e}_y$. Discrete translations by the cholesteric pitch $2\pi/q_0$ leave the structure invariant but arbitrary rotations about the pitch axis do not; instead there is screw symmetry under the combination of a rotation by angle $q_0 u$ and a translation by $u$, for any $u \in \mathbb{R}$. An equivalent way of expressing this, which emphasises the rotation-translation coupling of cholesteric hydrodynamics, is that a rotation by angle $q_0 u$ is equivalent to a translation by $-u$.

This rotation-translation coupling generates a chiral curl force even in the hydrodynamics of passive cholesterics~\cite{lubensky1972}, which distinguishes their chiral mechanics from two-dimensional odd elasticity, say, where Maxwell-Betti reciprocity prevents the odd forces from arising from a free energy~\cite{fruchart2023}. This aspect of cholesterics may be compared with the class of Hamiltonian curl forces described by Berry \& Shukla\footnote{We remark that their use of the term `curl force' is different from our usage here.}~\cite{berry2015}. Conceptually, the passive curl force can be understood from the underlying director theory for the cholesteric liquid crystal: Euclidean covariance entails using the co-rotational derivative for the director dynamics and the rotational coupling with fluid vorticity comes together with a corresponding antisymmetric term in the stress. Written in terms of the pseudolayer structure this stress is $\frac{1}{2q_0} h \,({\bf n}_{\perp} {\bf n} - {\bf n} {\bf n}_{\perp})$, where ${\bf n}$ is the director, ${\bf n}_{\perp} = {\bf N} \times {\bf n}$ and $h$ is the molecular field (of the pseudolayer displacement). Its divergence gives the curl force $\frac{1}{2q_0} \nabla \times ( h \,{\bf N} )$~\cite{lubensky1972}; we review its derivation in the next section. 

In active cholesterics, the chiral curl force $\zeta_c \nabla \times ( \nabla^2 u \,{\bf N} )$ generates vorticity about the pitch axis, which the rotation-translation coupling transforms into pseudolayer displacement. When the direction of the active chiral curl force is the same as that of the passive term its effects are stabilising. However, when the direction is opposite, sufficiently strong activity leads to a new form of active instability that is unique to the screw symmetry of cholesterics and sensitive to the handedness of the helical structure.

\section{Director Theory of Active Cholesterics}
\label{sec:cholesteric}

In this section we summarise and review the hydrodynamic theory of active liquid crystals with focus on its application to cholesterics. The hydrodynamics may be presented in terms of the director field or the Q-tensor; for simplicity, and to emphasise the geometric aspects, we adopt the director formalism. In a simplified presentation, the Ericksen-Leslie equations for the director field ${\bf n}$ and fluid velocity ${\bf v}$ can be given as incompressibility $\nabla \cdot {\bf v} = 0$ and force balance $\nabla \cdot \boldsymbol{\sigma} = 0$, together with an expression for the stress 
\begin{equation}
    \begin{split}
        \boldsymbol{\sigma} & = - p \,{\bf I} + 2 \eta \,{\bf D} - \frac{\partial f}{\partial \nabla {\bf n}} \cdot \bigl( \nabla {\bf n} \bigr)^T \\
        & \quad + \frac{1}{2} \bigl[ {\bf hn} - {\bf nh} \bigr] + \frac{\nu}{2} \bigl[ {\bf hn} + {\bf nh} \bigr] + \boldsymbol{\sigma}^{\textrm{a}} ,
    \end{split} \label{eq:EL_stress}
\end{equation}
and relaxational dynamics for the director field 
\begin{equation}
        \partial_t {\bf n} + {\bf v} \cdot \nabla {\bf n} + \boldsymbol{\Omega} \cdot {\bf n} = \frac{1}{\gamma} {\bf h} - \nu \Bigl[ {\bf D} \cdot {\bf n} - {\bf n} \bigl( {\bf D} : {\bf nn} \bigr) \Bigr] . 
    \label{eq:EL_n}
\end{equation}
Here, ${\bf D}$ and $\boldsymbol{\Omega}$ are the symmetric and antisymmetric parts of the velocity gradients, $\eta$ is an isotropic viscosity, $p$ is the pressure, $f$ is the free energy density, ${\bf h} = - \delta F/\delta {\bf n}$ is the molecular field, $\gamma$ is a rotational viscosity for the director relaxation, $\nu$ is the flow alignment parameter, and $\boldsymbol{\sigma}^{\textrm{a}}$ is the active stress. We take the active stress to have two contributions 
\begin{equation}
    \boldsymbol{\sigma}^{\textrm{a}} = - \zeta \biggl( {\bf nn} - \frac{1}{3} {\bf I} \biggr) - \zeta_c \Bigl( \nabla \times ({\bf nn}) + \bigl[ \nabla \times ({\bf nn}) \bigr]^T \Bigr) .
\end{equation}
The first is the usual force dipole term~\cite{Simha_Ramaswamy} with strength $\zeta$ that is positive in extensile systems. The second is the leading order chiral contribution and can be associated with the action of microscopic torque dipoles~\cite{furthauer2012,furthauer2013,markovich2019, Cates_drop,maitra2025}. We now specialise these equations to the hydrodynamics of cholesterics. 

Cholesterics are chiral nematics in which the director has an energetic preference for uniform twist~\cite{deGennesProst,ChaikinLubensky}. For simplicity we take the free energy to have the one-elastic-constant form  
\begin{equation}
    F = \int \frac{K}{2} \bigl| \nabla {\bf n} \bigr|^2 + K q_0 \,{\bf n} \cdot \nabla \times {\bf n} \,dV ,
    \label{eq:Frank}
\end{equation}
where $K$ is the elastic constant and $q_0$ is the chirality that determines the handedness and pitch of the cholesteric; the pitch is $2\pi/|q_0|$ and the cholesteric is right-handed for $q_0 > 0$. The ground state is a one-dimensional helical rotation of the director along one spatial direction, called the pitch axis, with the director lying everywhere perpendicular to this, for example ${\bf n} = \cos q_0z \,{\bf e}_x + \sin q_0z \,{\bf e}_y$ or any equivalent to it by a Euclidean motion. 

The hydrodynamics of cholesterics entails modulations of this structure on scales that are large compared to the pitch, $2\pi/|q_0|$. In the passive case, this was first determined by Lubensky~\cite{lubensky1972} and later revisited by Radzihovsky \& Lubensky~\cite{radzihovsky2011}, whose approach we adapt here. It consists of expressing the hydrodynamic modulation in terms of a helical phase field $\phi$ together with a moving frame ${\bf e}_i$ adapted to it, in which ${\bf e}_3 = \nabla\phi / |\nabla\phi| \equiv {\bf N}$ is the normal to the cholesteric pseudolayers (level sets of $\phi$) and the local pitch axis. This is illustrated in Fig.~\ref{fig:smectic_cholesteric}. The hydrodynamic variable is the displacement field of the pseudolayers, while the heliconical tilt of the director along the pitch axis is suppressed on scales larger than the pitch and, is therefore, non-hydrodynamic. The screw symmetry kinematically connects director rotations about the pitch axis with pseudolayer displacement and is manifested by the coupling of vorticity to the phase field dynamics and a corresponding antisymmetric contribution to the stress~\cite{lubensky1972}. We will show that this leads to a transverse flow response (in a passive cholesteric) analogous to the `odd' response of chiral activity~\cite{kole2021}. Further, the vorticity term in the pseudolayer hydrodynamics combines with chiral activity to alter the undulational Helfrich-Hurault instability of the dipolar activity~\cite{adhyapak2013,whitfield2017}, shifting both the threshold and selected wavevector by an amount depending on the handedness of both the chiral activity and the cholesteric helix. This effect derives from the screw symmetry in cholesterics and is absent in chiral smectics~\cite{kole2021}. In materials where mechanical stress compensates the dipolar active stress, chiral activity leads to instability and breakdown of cholesteric structure when its magnitude exceeds that of the twist contribution to director stresses. 

We show how the pseudolayer hydrodynamics can be obtained from the Ericksen-Leslie equations for the cholesteric director, retaining a geometric description of the cholesteric structure that can be applied more generally than a linearisation around the ground state. In terms of the helical phase field $\phi$ and moving frame ${\bf e}_i$, an entirely general director field can be written 
\begin{equation}
    {\bf n} = \cos\theta \bigl[ \cos q_0\phi \,{\bf e}_1 + \sin q_0\phi \,{\bf e}_2 \big] + \sin\theta \,{\bf e}_3 ,
    \label{eq:hydro_n}
\end{equation}
where $\theta$ is a heliconical angle. We denote the connection 1-forms for the frame by $\omega^j_i = (\nabla {\bf e}_i) \cdot {\bf e}_j$ and their components by $\omega^j_{ki} = \omega^j_i({\bf e}_k)$; the integrability of the plane spanned by ${\bf e}_1, {\bf e}_2$ yields the equality $\omega^3_{12} = \omega^3_{21}$. They are constrained by Cartan's equations of structure $d\omega^i_j + \omega^i_k \wedge \omega^k_j = 0$, since $\mathbb{R}^3$ is flat; we will only make use of the linearised form of this condition, which is that the connection forms are closed and therefore exact. The connection forms $\omega^3_i = - \nabla {\bf N} \cdot {\bf e}_i$ are determined (almost) directly by the phase field and are essentially geometric. In contrast, $\omega^2_1$ is not purely geometric and requires some analysis to identify; we will see that it is connected to the bending elasticity and that its sign depends on the handedness of the cholesteric (sign of $q_0$). 

We present this technical part of the analysis first. It follows from the hydrodynamic form of the molecular field, which we express using a frame adapted to the director and the plane orthogonal to it~\cite{machon2016PRX,pollard2021}, for which we introduce the basis (${\bf n} \equiv {\bf E}_1$) 
\begin{align}
    {\bf E}_2 & = - \sin q_0\phi \,{\bf e}_1 + \cos q_0\phi \,{\bf e}_2 , \\
    {\bf E}_3 & = - \sin\theta \bigl( \cos q_0\phi \,{\bf e}_1 + \sin q_0\phi \,{\bf e}_2 \bigr) + \cos\theta \,{\bf N} .
\end{align}
In this basis the molecular field is 
\begin{equation}
    \begin{split}
        {\bf h} & = K \cos\theta \Bigl[ q_0 \bigl( \nabla^2\phi + \omega^3_{11} + \omega^3_{22} \bigr) + \nabla \cdot \omega^2_1 + \cdots \Bigr] \,{\bf E}_2 \\
        & \quad + K \biggl( \nabla^2 \theta - q_0^2 \sin\theta \cos\theta \\
        & \qquad + \cos q_0\phi \Bigl[ \nabla \cdot \omega^3_1 - 2q_0 \omega^2_{11} + 2q_0 \omega^3_{32} \Bigr] \\
         & \quad + \sin q_0\phi \Bigl[ \nabla \cdot \omega^3_2 - 2q_0 \omega^2_{21} - 2q_0 \omega^3_{31} \Bigr] + \cdots \biggr) {\bf E}_3 ,
    \end{split}        
\end{equation}
where we have suppressed subdominant terms. The ${\bf E}_3$-component controls the heliconical angle; its leading order linear part near the Brillouin zone centre is $\nabla^2 \theta - q_0^2 \theta$ and implies $\theta$ is gapped on scales large compared to the cholesteric pitch. Continuing a multi-scale analysis, the non-hydrodynamic part of $\theta$ can be expanded as a Fourier series in $\cos nq_0\phi$ and $\sin nq_0\phi$ with coefficients that are slowly varying. Consideration of the leading order balance for the $n=1$ modes gives 
\begin{gather}
    \omega^2_{i1} = \frac{1}{2q_0} \nabla \cdot \omega^3_i , \quad i = 1,2, \\ 
    \theta = \frac{1}{q_0} \omega^3_{32} \cos q_0\phi - \frac{1}{q_0} \omega^3_{31} \sin q_0 \phi .
\end{gather}
For the linearisation about the cholesteric ground state, $\phi = z - u$ and $\omega^3_i = \nabla \partial_i u$, so this identifies the connection form $\omega^2_1$ with $\frac{1}{2q_0} \nabla (\nabla^2 u)$. In this linearised form it is the gradient of the pseudolayer mean curvature divided by $q_0$; the gradient of the mean curvature is the normal stress of a Canham-Helfrich membrane~\cite{capovilla2002,al-izzi2023}. 

From now on we retain only the hydrodynamic part of the director field, which lies entirely in the tangent plane to the helical phase field ($\theta = 0$), and the hydrodynamic part of the molecular field, which is  
\begin{equation}
    {\bf h} = K \Bigl[ q_0 \bigl( \nabla^2 \phi + \omega^3_{11} + \omega^3_{22} \bigr) + \nabla \cdot \omega^2_1 \Bigr] {\bf E}_2 . 
\label{eq:cholesteric_hydro_h}
\end{equation}
Considering first the linearisation about the cholesteric ground state ($\phi = z - u$), we find $\nabla^2 \phi + \omega^3_{11} + \omega^3_{22} = - \nabla^2 u + \partial_{xx} u + \partial_{yy} u = - \partial_{zz} u$ and $\nabla \cdot \omega^2_1 = \frac{1}{2q_0} \nabla^4 u$, so that the 1-forms $\omega^3_i$ serve to eliminate the `layer tilt' term $\nabla_{\perp}^2 u$ from $\nabla^2 \phi$ and give the usual linearised compressional elasticity, while the bending elasticity comes from the connection 1-form $\omega^2_1$. A purely geometric description can be given for the compressional elasticity, since the combination $\omega^3_{11} + \omega^3_{22}$ is twice the pseudolayer mean curvature (it is the trace of the shape operator for the pseudolayers). The way the connection forms enter into the molecular field and their relation to the pseudolayer elasticity can be contrasted with how they enter into the hydrodynamic analysis of the free energy~\cite{radzihovsky2011}; there the forms $\omega^3_1$, $\omega^3_2$ contribute the bending elasticity and $\omega^2_1$ can be neglected in the leading order analysis. 

We now deduce the hydrodynamics of the phase field from the Ericksen-Leslie equations for the director field and fluid flow, using the hydrodynamic form of the molecular field~\eqref{eq:cholesteric_hydro_h}. To extract the equation for the phase field we project the relaxation equation for the director~\eqref{eq:EL_n} on the direction ${\bf E}_{2}$, which gives 
\begin{equation}
    \begin{split}
        & q_0 \partial_t \phi + \partial_t {\bf e}_1 \cdot {\bf e}_2 + {\bf v} \cdot \nabla {\bf n} \cdot {\bf E}_{2} + \boldsymbol{\Omega} : {\bf E}_{2} {\bf n} \\
        & \quad = \frac{1}{\gamma} \,{\bf h} \cdot {\bf E}_{2} - \nu \,{\bf D} : {\bf E}_{2} {\bf n} .
    \end{split}
\end{equation}
Our focus is on the hydrodynamic part of this equation that varies much more slowly than the scale of the cholesteric pitch. To this end, we write 
\begin{equation}
    \begin{split}
        {\bf E}_{2} {\bf n} & = \frac{1}{2} \bigl[ {\bf e}_2 {\bf e}_1 - {\bf e}_1 {\bf e}_2 \bigr] - \frac{1}{2} \sin 2q_0\phi \;\bigl[ {\bf e}_1 {\bf e}_1 - {\bf e}_2 {\bf e}_2 \bigr] \\
        & \quad + \frac{1}{2} \cos 2q_0\phi \bigl[ {\bf e}_1 {\bf e}_2 + {\bf e}_2 {\bf e}_1 \bigr] ,
    \end{split}
    \label{eq:chol_hydro_E2n}
\end{equation}
and subsequently neglect the last two terms on the right-hand side. This yields the hydrodynamic equation for the pseudolayer phase field  
\begin{equation}
    \begin{split}
        & \partial_t \phi + {\bf v} \cdot \biggl( \nabla \phi + \frac{1}{q_0} \omega^2_1 \biggr) - \frac{1}{2q_0} \bigl( {\bf N} \cdot \nabla \times {\bf v} \bigr) \\
        & \quad = \frac{K}{\gamma} \biggl( \nabla^2 \phi + \omega^3_{11} + \omega^3_{22} + \frac{1}{q_0} \nabla \cdot \omega^2_1 \biggr) ,
    \end{split}
    \label{eq:chol_hydro}
\end{equation}
where we have dropped a term $\frac{1}{q_0} \partial_t {\bf e}_1 \cdot {\bf e}_2$ coming from the time-dependence of the local frame as being higher-order in spatial derivatives\footnote{Consider linear perturbations around the cholesteric ground state. The locally adapted frame of the perturbed structure can be expanded in terms of the basis of the ground state:
\begin{equation*}
    {\bf e}_1 = {\bf e}_x + \theta_{12} \,{\bf e}_y + \theta_{13} \,{\bf e}_z ,
    \end{equation*}
    \begin{equation*}
    {\bf e}_2 = {\bf e}_y - \theta_{12} \,{\bf e}_x + \theta_{23} \,{\bf e}_z , 
    \end{equation*}
    \begin{equation*}
    {\bf e}_3 = {\bf e}_z - \theta_{13} \,{\bf e}_x - \theta_{23} \,{\bf e}_y ,
    \end{equation*}
where the $\theta_{ij}$ are all small. The term in question is then 
\begin{equation*}
\frac{1}{q_0} \partial_t {\bf e}_1 \cdot {\bf e}_2 = \partial_t \theta_{12} + \textrm{non-linear}.
\end{equation*}
 At the same time, the connection form $\omega^2_1$ is $\omega^2_1 = \nabla {\bf e}_1 \cdot {\bf e}_2 = \nabla \theta_{12} + \textrm{non-linear}$, and, since $\omega^2_1 = \frac{1}{2q_0} \nabla ( \nabla^2 u )$ at linear order, we can conclude that 
\begin{equation*}
    \frac{1}{q_0} \partial_t {\bf e}_1 \cdot {\bf e}_2 = \frac{1}{2q_0^2} \nabla^2 \partial_t u.
\end{equation*}
This is smaller than $\partial_t \phi = - \partial_t u$ by a factor of order $\nabla^2$, which motivates our neglect of it.} than the leading term $\partial_t \phi$. The key feature of screw symmetry in cholesterics is the coupling of fluid vorticity parallel to the pitch axis with the phase field dynamics. Linearising about the cholesteric ground state, $\phi = z - u$, we find~\eqref{eq:chol_hydro} reduces to 
\begin{equation}
    \partial_t u - v_z + \frac{1}{2q_0} \bigl( {\bf e}_z \cdot \nabla \times {\bf v} \bigr) = \frac{K}{\gamma} \biggl( \partial_{zz} u - \frac{1}{2q_0^2} \,\nabla^4 u \biggr) .
    \label{eq:ueq}
\end{equation}
The bulk modulus is $B = K q_0^2$ and the bending modulus is $\bar{K} = K/2$, in agreement with~\cite{radzihovsky2011}. 

Turning now to the Stokes equation, we first calculate the term coming from the Ericksen stress. We use the identity $\nabla \cdot [(\partial f / \partial \nabla {\bf n}) \cdot (\nabla {\bf n})^T] = \nabla {\bf n} \cdot {\bf h} + \nabla f$ and absorb the $\nabla f$ term into the pressure~\cite{deGennesProst}, leaving 
\begin{equation}
    \begin{split}
        \nabla {\bf n} \cdot {\bf h} & = K q_0^2 \biggl( \nabla^2 \phi + \omega^3_{11} + \omega^3_{22} + \frac{1}{q_0} \nabla \cdot \omega^2_1 \biggr) \\
        & \hspace{10mm} \times \biggl( \nabla \phi + \frac{1}{q_0} \omega^2_1 \biggr) ,
    \end{split}
\end{equation}
neglecting all non-hydrodynamic terms. For the remaining nematic stress terms we take the hydrodynamic part of ${\bf hn}$ in the same way as~\eqref{eq:chol_hydro_E2n}, which shows that the symmetric term $\frac{\nu}{2} \bigl( {\bf hn} + {\bf nh} \bigr)$ is non-hydrodynamic. It is interesting to note that the hydrodynamics of cholesterics comes from the Ericksen stress and not the flow alignment; by contrast, in nematics the flow alignment is important to the linear hydrodynamics, including the active instability, and the Ericksen stress only enters at non-linear order. As in smectic hydrodynamics this arises because advection enters the linearised displacement field dynamics and the Ericksen stress is its Onsager counterpart. The difference in cholesterics, as we have mentioned, is that the vorticity also enters the displacement field dynamics and so its Onsager counterpart, the antisymmetric stress, remains hydrodynamic, albeit higher order in gradients. The hydrodynamic part of the Stokes equation is therefore 
\begin{widetext}
\begin{equation}
\begin{split}
    0 & = - \nabla p + \eta \nabla^2 {\bf v} - K q_0^2 \biggl( \nabla^2 \phi + \omega^3_{11} + \omega^3_{22} + \frac{1}{q_0} \nabla \cdot \omega^2_1 \biggr) \biggl( \nabla \phi + \frac{1}{q_0} \omega^2_1 \biggr) \\
    & \quad + \frac{Kq_0}{2} \,\nabla \times \biggl( \nabla^2 \phi + \omega^3_{11} + \omega^3_{22} + \frac{1}{q_0} \nabla \cdot \omega^2_1 \biggr) {\bf N} + \nabla \cdot \boldsymbol{\sigma}^{\textrm{a}} .
\end{split}
\label{eq:chol_hydro_Stokes}
\end{equation}
\end{widetext}
Taken together, \eqref{eq:chol_hydro} and~\eqref{eq:chol_hydro_Stokes} provide a hydrodynamic description of cholesterics that retains full geometric non-linearity with the only assumption being the hydrodynamic form of the director~\eqref{eq:hydro_n}. As a result, they can be applied much more generally than just to states close to the helical ground state, although we do not pursue this explicitly here. A feature of cholesteric hydrodynamics is the presence of a chiral curl force $\frac{1}{2q_0} \nabla \times (h {\bf N})$, where $h$ is the (hydrodynamic) molecular field of the pseudolayer structure, which is the required Onsager anti-symmetric (reactive) counterpart to the kinematic term $(1/2q_0) \,{\bf N}\cdot\nabla\times{\bf v}$ in~\eqref{eq:chol_hydro}. This derives from the screw symmetry of cholesterics and distinguishes their equilibrium hydrodynamics from those of smectics. To our knowledge, it is the only currently known example of an odd mechanical force in equilibrium elasticity, although it is clear from its origin that similar odd forces will arise for any mesophase with screw symmetry, {\sl e.g.} the blue phases.     

Finally, we present the hydrodynamic form of the active stress in a cholesteric, which is a much simpler calculation. We use the director field~\eqref{eq:hydro_n} and the identity 
\begin{equation}
    \begin{split}
        {\bf nn} & = \frac{1}{2} \bigl[ {\bf e}_1 {\bf e}_1 + {\bf e}_2 {\bf e}_2 \bigr] + \frac{1}{2} \cos 2q_0\phi \bigl[ {\bf e}_1 {\bf e}_1 - {\bf e}_2 {\bf e}_2 \bigr] \\
        & \quad + \frac{1}{2} \sin 2q_0\phi \bigl[ {\bf e}_1 {\bf e}_2 + {\bf e}_2 {\bf e}_1 \bigr] ,
    \end{split}
\end{equation}
to obtain the hydrodynamic form  
\begin{equation}
    \begin{split}
        \boldsymbol{\sigma}^{\textrm{a}} & = \frac{\zeta}{2} \biggl( {\bf NN} - \frac{1}{3} {\bf I} \biggr) \\
        & \quad + \frac{\zeta_c}{2} \Bigl( \nabla \times ({\bf NN}) + \bigl[ \nabla \times ({\bf NN}) \bigr]^T \Bigr) , 
    \end{split}
    \label{eq:active_stress_hydro}
\end{equation}
discarding the non-hydrodynamic terms. The interpretation is that the active stress in a cholesteric has the same hydrodynamic form as the active stress in a (chiral) smectic A, with the correspondence being that an extensile cholesteric is equivalent to a contractile smectic~\cite{whitfield2017,kole2021}; similarly there is a change in sign for the chiral active stress. The active force density arising from the hydrodynamic active stress~\eqref{eq:active_stress_hydro} is 
\begin{equation}
    \begin{split}
        \nabla \cdot \boldsymbol{\sigma}^{\textrm{a}} & = \frac{\zeta}{2} \Bigl[ {\bf N} \bigl( \nabla \cdot {\bf N} \bigr) + \bigl( {\bf N} \cdot \nabla \bigr) {\bf N} \Bigr] \\
        & \quad + \frac{\zeta_c}{2} \nabla \times \Bigl[ {\bf N} \bigl( \nabla \cdot {\bf N} \bigr) + \bigl( {\bf N} \cdot \nabla \bigr) {\bf N} \Bigr] .
    \end{split}
\end{equation}
The chiral contribution coming from the torque-dipole stress is a curl force that represents an odd, or chiral, mechanical response. In cholesterics, bend in the pitch axis is suppressed~\cite{lubensky1972}, so that the dominant contribution of this curl force is $- \frac{\zeta_c}{2} \,{\bf N} \times \nabla ( \nabla \cdot {\bf N} ) \simeq - \zeta_c \,{\bf N} \times \nabla H$, where $H$ is the pseudolayer mean curvature~\cite{kole2021}.

\subsection{Mechanical Strain}
\label{subsec:mechanical_strain}

The achiral active stress is analogous to that induced in an equilibrium cholesteric by an imposed mechanical strain~\cite{kole2021}. We comment here on how an imposed strain affects the equilibrium curl force in cholesterics. 

A state of planar pseudolayers dilated by a strain $\alpha$ is described by the phase field $\phi = \frac{1}{1+\alpha} \,z$. Linear fluctuations around this state can be encoded either by writing $\phi = \frac{1}{1+\alpha} (z - u)$ or $\phi = \frac{1}{1+\alpha} \,z - u$. Adopting the former, the linearised pseudolayer normal is ${\bf N} = {\bf e}_z - \nabla_{\perp} u$, the same expression as for the unstrained layers ($\alpha = 0$) and the connection 1-forms are then also given by the same expressions as before. However, $\nabla^2 \phi = - \frac{1}{1+\alpha} \nabla^2 u$ so that 
\begin{multline}
    \nabla^2 \phi + \omega^3_{11} + \omega^3_{22} + \frac{1}{q_0} \nabla \cdot \omega^2_1 =\\ - \frac{1}{1+\alpha} \,\partial_{zz} u + \frac{\alpha}{1+\alpha} \nabla_{\perp}^2 u + \frac{1}{2q_0^2} \nabla^4 u + \cdots .
\end{multline}
To leading order, the strain produces an additional term proportional to $\nabla_{\perp}^2 u$. The consequence for the chiral force in~\eqref{eq:chol_hydro_Stokes} is to generate a term of the form $\frac{Kq_0}{2} \frac{\alpha}{1+\alpha} \,{\bf e}_z \times \nabla \nabla_{\perp}^2 u$, with the same essential structure as arises from the chiral active stress. The difference is that, in the case of mechanical strain, the term appears consistently everywhere that the pseudolayer molecular field enters the equations so that there is still an equilibrium solution in which the molecular field vanishes and there is no flow. 

Note also that, although the static pseudolayer structure is the same as that of a smectic, the dynamic evolution to this state is different. This is because the curl force couples pseudolayer motion along its normal to fluid vorticity about it. For example, the dynamics of the Helfrich-Hurault undulational instability involves a vortex lattice of fluid flow. This makes sense as the motion of the pseudolayers in a cholesteric is equivalent to the rotation of the molecules within the tangent planes.

\section{Nonlinear Dynamical Equations for the Displacement Field}
\label{sec:nonlinear}

In this section, we discuss the nonlinear displacement field dynamics obtained from the director theory of active cholesterics. The dynamics of the displacement field is 
\begin{equation}
    \partial_tu+{\bf v}\cdot\nabla u+\frac{1}{2q_0}({\bf N}\cdot\nabla\times{\bf v})=v_z-\frac{1}{\gamma q_0^2}\frac{\delta F_u}{\delta u}\,,
\end{equation}
where the free energy $F_u$ is displayed below. For the active force densities, it is useful to define a polar vector whose integral is conserved and which has nematic symmetry, i.e., it is invariant under director inversion (and inversion of the layer normal):
\begin{multline}
    {\bf P}=-\frac{1}{|\nabla\phi|^2}\left[\nabla\phi\left(\frac{\nabla\phi\cdot\nabla|\nabla\phi|^2}{|\nabla\phi|^2}+\nabla^2 u\right)+\nabla\phi\cdot\nabla\nabla u\right]\\=\Bigl[ {\bf N} \bigl( \nabla \cdot {\bf N} \bigr) + \bigl( {\bf N} \cdot \nabla \bigr) {\bf N} \Bigr]\,.
\end{multline}
With this definition, the force balance equation reads
\begin{equation}
\label{frcbalance}
    \eta\nabla^2{\bf v}=\nabla p+\frac{\delta F_u}{\delta u}\nabla\phi+\frac{1}{2q_0}{\bf e}_z\times\nabla\frac{\delta F_u}{\delta u}-\frac{\zeta}{2}{\bf P}-\frac{\zeta_c}{2}\nabla\times{\bf P}\,.
\end{equation}
Neglecting the heliconical angle $\theta$ {\sl a priori} the hydrodynamic part of the free energy~\eqref{eq:Frank} is 
\begin{equation}
\begin{split}
    F & = \int - \frac{Kq_0^2}{2} + \frac{Kq_0^2}{2} \biggl( |\nabla\phi| - 1 + \frac{1}{q_0} \omega^2_{31} \biggr)^2 \\
    & \quad + \frac{K}{2} \Bigl( |\omega^2_1|^2 - \bigl( \omega^2_{31} \bigr)^2 \Bigr) + \frac{K}{4} \Bigl( |\omega^3_1|^2 + |\omega^3_2|^2 \Bigr) \,dV .
\end{split}
\end{equation}
Dropping certain naturally subleading terms, and the constant $-Kq_0^2/2$, this reduces to the rotationally-invariant smectic free energy 
\begin{equation}
    F_u=\frac{Kq_0^2}{2}\int\left\{\left[\partial_z u-\frac{(\nabla u)^2}{2}\right]^2+\frac{1}{2q_0^2}(\nabla^2 u)^2\right\}dV\,,
\end{equation}
for the Eulerian displacement field. Notice that the form of the active force densities in \eqref{frcbalance} and the one discussed in \cite{kole2021} are different. In that work, the active force densities were constructed from the gradients of a mass-density wave and not from the director field, as we do here. As we discuss in that work, this difference in the achiral active force density can be absorbed in a redefinition of the free energy $F_u$ that appears in the force balance equation, without a corresponding redefinition in the displacement field equation. Since the dynamics of an active system is not controlled by a free energy, even the force densities that have the same structure as terms in equilibrium need not arise as a functional derivative of the potential that controls permeative flow\footnote{A similar issue was discussed for active nematics in \cite{Green_Toner_Vitelli}; moreover, even the standard active stress in nematics \cite{Simha_Ramaswamy} can be viewed as arising from a disagreement between a free energy-like functional that controls force densities and the one that controls the dynamics of the order parameter, with the two having distinct critical points.}. This additional active force density did not affect the discussion in \cite{kole2021}, which concerned itself with the physics at leading order in wavenumbers: the effect of permeation appeared only at higher order in gradients and, therefore, the disagreement between the free energies was not relevant.

The distinction between the chiral active force density in this work and in \cite{kole2021} can also be absorbed in a redefinition of $F_u$ in the term $\propto{\bf e}_z\times\nabla(\delta F_u/\delta u)$. This is not the same renormalisation as in the last paragraph: in an active system, the ``free energies'' from which various force densities are obtained need not agree. Interestingly, in systems that are not {\sl structurally} chiral (i.e., when $q_0=0$) but are still composed of chiral elements, such as active smectic A* or cholesteric liquid crystals at the compensation point, there is no passive force density $\propto{\bf e}_z\times\nabla(\delta F_u/\delta u)$. This implies that in those materials, the distinction between the chiral active force density in \cite{kole2021} and the one here cannot be absorbed in a redefinition of the energy: there is no equilibrium term with an equivalent gradient structure. Indeed, neither this work nor \cite{kole2021} examined the most general form of the active (chiral and achiral) force densities in layered systems. Such systems break three-dimensional rotation symmetry by spontaneously choosing the layering direction (or the pitch direction in the case of cholesterics). Therefore, even at linear order and to lowest order in gradients, the most general active achiral force density should have the form $(\zeta_\parallel\partial_{zz}u+\zeta_\perp\nabla_\perp^2 u){\bf e}_z$, apart from terms that are a pure gradient and can be absorbed in a redefinition of the pressure. Similarly, the chiral active force density should have the form ${\bf e}_z\times\nabla(\zeta_{c_\parallel}\partial_{zz}u+\zeta_{c_\perp}\nabla_\perp^2 u)$ \cite{kole2024}. The active force densities used in this section (and therefore, the linearised one in the next section) have a particular relation between $\zeta_\perp$ and $\zeta_\parallel$ ($\zeta_\parallel/\zeta_\perp=-1$) and $\zeta_{c_\perp}$ and $\zeta_{c_\parallel}$ ($\zeta_{c_\parallel}/\zeta_{c_\perp}=-1$) for simplicity, which is not generic and not demanded by any symmetry; this relation need not be preserved under renormalisation.

\section{Linearised Hydrodynamics of Active Cholesterics}
\label{sec:linear_cholesteric}

We now describe the consequences of the screw symmetry in cholesterics for the linear stability of the helical ground state, with focus on the terms expressing the rotation-translation coupling. In part their effects are higher order in wavenumber compared to those of the usual dipolar active stress, however, the correspondence between dipolar activity and mechanical strain also allows for an additional instability at the compensation point. 

The linearisation of~\eqref{eq:chol_hydro} and~\eqref{eq:chol_hydro_Stokes} around the helical ground state, $\phi = z - u$, yields 
\begin{widetext}
    \begin{gather}
        \partial_t u - v_z + \frac{1}{2q_0} \bigl( {\bf e}_z \cdot \nabla \times {\bf v} \bigr) = \frac{K}{\gamma} \biggl( \partial_{zz} u - \frac{1}{2q_0^2} \nabla^4 u \biggr) , \label{eq:linear_udynamics} \\
        \begin{split}
            0 & = - \nabla p + \eta \nabla^2 {\bf v} + K q_0^2 \biggl( \partial_{zz} u - \frac{1}{2q_0^2} \nabla^4 u \biggr) {\bf e}_z + \frac{Kq_0}{2} \,{\bf e}_z \times \nabla \biggl( \partial_{zz} u - \frac{1}{2q_0^2} \nabla^4 u \biggr) \\
            & \quad - \frac{\zeta}{2} \Bigl( {\bf e}_z \nabla_{\perp}^2 u + \nabla_{\perp} \partial_z u \Bigr) + \frac{\zeta_c}{2} \,{\bf e}_z \times \nabla \Bigl( \nabla_{\perp}^2 u - \partial_{zz} u \Bigr) .
        \end{split} \label{eq:linear_Stokes}
    \end{gather}
\end{widetext}
We solve the Stokes equation by Fourier transforming it. By incompressibility the fluid flow is orthogonal to the wavevector ${\bf q}$ of the Fourier mode and we introduce an adapted orthonormal frame $\{{\bf e}_1=\hat{{\bf q}}, {\bf e}_2, {\bf e}_3\}$ to describe it. We take ${\bf e}_3$ along the intersection of the plane orthogonal to ${\bf q}$ with the $xy$-plane orthogonal to the (unperturbed) pseudolayer normal. Then, writing ${\bf q} = q_z \,{\bf e}_z + {\bf q}_{\perp}$ we have ${\bf e}_2 = - (q_{\perp}/q) \,{\bf e}_z + (q_z/q) \,\hat{{\bf q}}_{\perp}$. We find the fluid velocity is given by 
\begin{equation}
    \begin{split}
        & {\bf v}_{{\bf q}} = {\bf e}_2 \biggl[ \frac{Kq_0^2 q_{\perp}}{\eta q^3} \biggl( q_z^2 + \frac{q^4}{2q_0^2} \biggr) - \frac{\zeta q_{\perp} (q_{\perp}^2 - q_z^2)}{2\eta q^3} \biggr] u_{\bf q} \\
        & \;\; - i \,{\bf e}_3 \biggl[ \frac{Kq_0 q_{\perp}}{2\eta q^2} \biggl( q_z^2 + \frac{q^4}{2q_0^2} \biggr) + \frac{\zeta_c q_{\perp} (q_{\perp}^2 - q_z^2)}{2\eta q^2} \biggr] u_{\bf q} .
    \end{split}
\end{equation}
The chiral, or odd, component is the part in the direction ${\bf e}_3$ and has both active and passive contributions. The passive contribution, originating in the screw symmetry of cholesterics, is proportional to $q_0$ and so its direction depends on the handedness of the cholesteric. 

\begin{figure*}[t!]
    \centering
    \includegraphics[width=\textwidth]{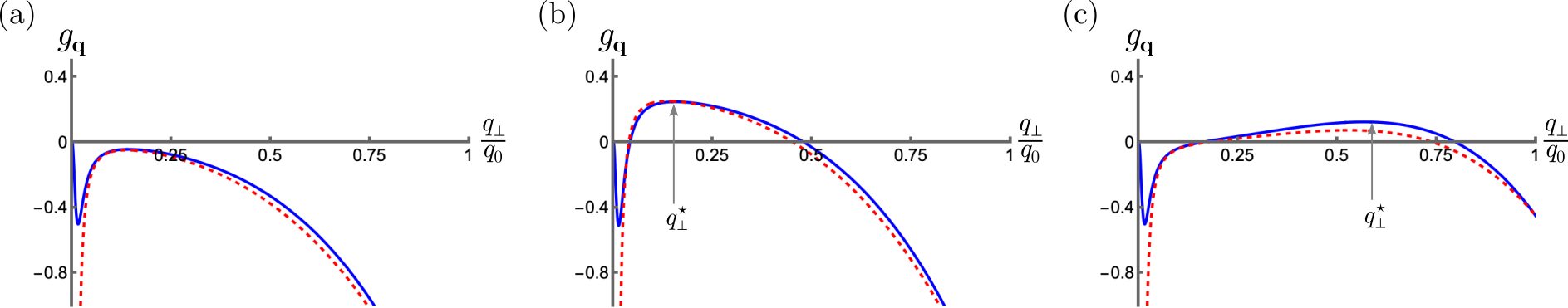}
    \caption{Linearised growth rate $g_{\bf q}$ (made dimensionless) and active instabilities in cholesterics. The solid blue line shows the full function~\eqref{eq:gq_full} and the dashed red line shows the approximation~\eqref{eq:gq_approx}. The minimum in the full function at small $q_{\perp}/q_0$ corresponds to the scale of confinement along the pitch axis, $q_{\perp} \simeq |q_z|$. (a) Passive cholesteric, $\zeta = \zeta_c = 0$. (b) Helfrich-Hurault-type instability for $\zeta > \zeta_{\textrm{th}}$; the most unstable wavevector $q_{\perp}^{\star}$ is indicated. (c) Chiral instability for $|\zeta_c| > |\zeta_c^{\textrm{th}}|$; the most unstable wavevector $q_{\perp}^{\star}$ is indicated.}
    \label{fig:growth}
\end{figure*}

In the linearised dynamics of the pseudolayer displacement field, the ${\bf e}_2$-component of flow enters through the advective coupling, while the ${\bf e}_3$-component contributes via the rotation-translation coupling of the cholesteric screw symmetry. As a consequence, the chiral activity $\zeta_c$ enters the linear displacement field dynamics in cholesterics whereas it does not in smectic-A. The resulting linearised dynamics is $\partial_t u_{\bf q} = g_{\bf q} u_{\bf q}$, with linearised growth rate 
\begin{equation}
    \begin{split}
        g_{\bf q} & = - \frac{K}{\eta} \biggl( \frac{q_0^2 q_{\perp}^2}{q^4} + \frac{q_{\perp}^2}{4q^2} + \frac{\eta}{\gamma} \biggr) \biggl( q_z^2 + \frac{q^4}{2q_0^2} \biggr) \\
        & \quad + \biggl( \frac{\zeta}{2\eta} - \frac{\zeta_c q^2}{4\eta q_0} \biggr) \frac{q_{\perp}^2 (q_{\perp}^2 - q_z^2)}{q^4} .
    \end{split}
    \label{eq:gq_full}
\end{equation}

This linear growth rate contains instabilities associated to both active stresses, $\zeta$ and $\zeta_c$. We show plots of the growth function illustrating these instabilities in Fig.~\ref{fig:growth}. The essential character of both, and all important properties, can be obtained from a simplified form of the growth rate. The fundamental hydrodynamic instability is associated to modes with $q_{\perp}^2 > q_z^2$ that are primarily undulational, rather than compressional, and in this regime we expand the growth rate in $q_z$, retaining only the most relevant terms 
\begin{equation}
    \begin{split}    
        g_{\bf q} & = \frac{\zeta}{2\eta} - \frac{K + \zeta_c / 2q_0}{2\eta} q_{\perp}^2 - \frac{K}{2\eta q_0^2} \biggl( \frac{1}{4} + \frac{\eta}{\gamma} \biggr) q_{\perp}^4 \\
        & \quad - \frac{Kq_0^2 q_z^2}{\eta q_{\perp}^2} .
    \end{split}
    \label{eq:gq_approx}
\end{equation}
The activity contributes to both terms of $O(q_{\perp}^0)$ and $O(q_{\perp}^2)$, and the nature of the instability depends on the sign of both terms. Our main focus here will be on the chiral activity $\zeta_c$ and the sign of $K+\zeta_c / 2q_0$; we begin with the case where it is positive, Fig.~\ref{fig:growth}(b). 

The term retained in $q_z^2$ serves to control the behaviour at small $q_{\perp}$ and is important when the system is confined by cell boundaries along the direction of the pitch axis (and consequently $|q_z|$ takes a minimum value $|q_z| \approx \pi/d$, where $d$ is the cell gap). Its effect is to select the most unstable wavevector for the Helfrich-Hurault-type undulational instability~\cite{adhyapak2013,whitfield2017,kole2021,TranRMP}. The critical points of $g_{\bf q}$ with respect to $q_{\perp}$ are given by 
\begin{equation}
    0 = - \frac{K + \zeta_c / 2q_0}{2\eta} - \frac{K}{\eta q_0^2} \biggl( \frac{1}{4} + \frac{\eta}{\gamma} \biggr) q_{\perp}^2 + \frac{Kq_0^2 q_z^2}{\eta q_{\perp}^4} .
\end{equation}
The dominant balance is between the first and last terms and identifies the most unstable transverse wavevector 
\begin{equation}
    q_{\perp}^{\star} = 2^{1/4} \sqrt{|q_0 q_z|} \,\biggl[ 1 + \frac{\zeta_c}{2Kq_0} \biggr]^{-1/4} .
\end{equation} 
The geometric mean scaling $q_{\perp}^{\star} \sim$ $\sqrt{|q_0 q_z|}$ is a hallmark of the Helfrich-Hurault-type instability~\cite{TranRMP}. Substituting this value into the growth rate $g_{\bf q}$ gives a threshold value of activity for instability 
\begin{equation}
    \zeta_{\textrm{th}} = 2 \sqrt{2} \,|q_0 q_z| K \biggl( 1 + \frac{\zeta_c}{2Kq_0} \biggr)^{1/2} ,
\end{equation}
with the usual scalings of the Helfrich-Hurault instability~\cite{TranRMP}. The main difference in cholesterics as compared to smectics is that both the threshold $\zeta_{\textrm{th}}$ and most unstable wavevector $q_{\perp}^{\star}$ have coefficients that depend on the strength of chiral activity $\zeta_c$ in conjunction with the handedness of the cholesteric (sign of $q_0$). 

Now we consider the case where $K + \zeta_c / 2q_0$ is negative, where the character of the instability is different, see Fig.~\ref{fig:growth}(c). In this case, the term in $q_z^2$ is not essential and for simplicity of presentation we omit it. The approximated linear growth rate is then a quartic polynomial in $q_{\perp}$ with a local minimum at $q_{\perp} = 0$ and a maximum at 
\begin{equation}
    q_{\perp}^{\star} = q_0 \,\biggl| \frac{K + \zeta_c / 2q_0}{2K} \biggr|^{1/2} \biggl( \frac{1}{4} + \frac{\eta}{\gamma} \biggr)^{-1/2} .
\end{equation}
The lengthscale of the undulation therefore scales with the cholesteric pitch and the instability is only hydrodynamic if the prefactor is small, {\sl i.e.} if $K + \zeta_c / 2q_0$ is not too negative. This suggests that sufficiently strong chiral activity, with opposite handedness to the cholesteric helix, will lead to breakdown of the cholesteric structure at the level of the pitch. The threshold for this instability is $\zeta_c^{\textrm{th}} = - 2Kq_0$ when $\zeta$ is positive or zero, while when $\zeta$ is negative it is given by 
\begin{equation}
    \zeta_c^{\textrm{th}} = - 2K q_0 - \frac{2q_0 \sqrt{-\zeta K}}{|q_0|} \biggl( 1 + \frac{4\eta}{\gamma} \biggr)^{1/2} .
\end{equation}

\begin{figure}[t]
    \centering
    \includegraphics[width=\linewidth]{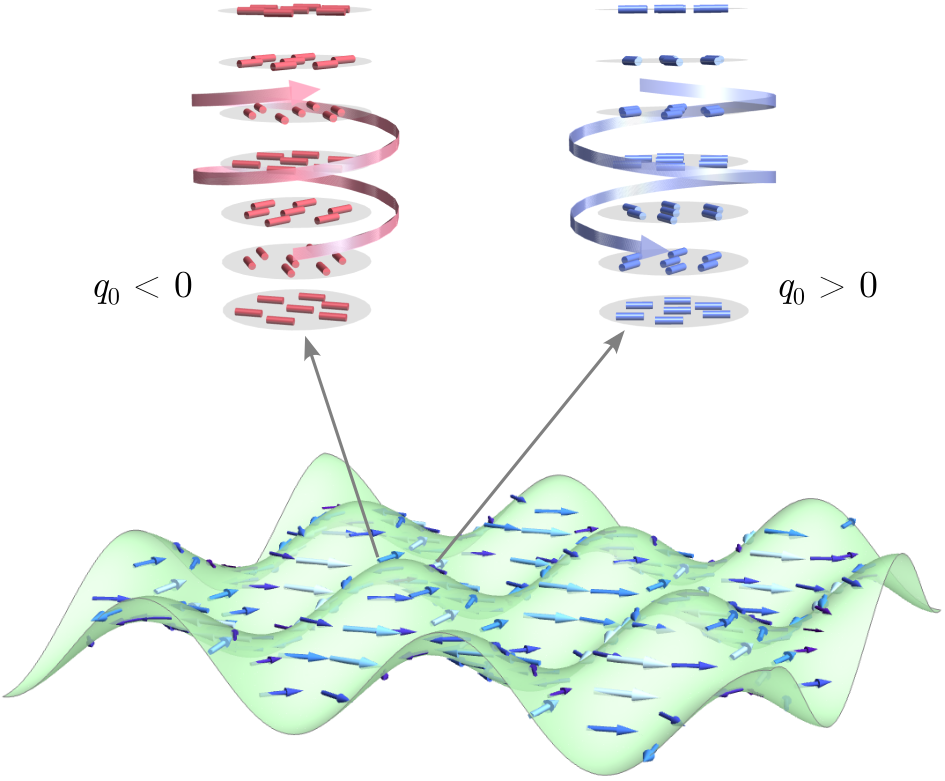}
    \caption{Odd active instability in cholesterics. Pseudolayer undulations, here a square lattice, are accompanied by in-layer vortical flows generated by the chiral activity, shown for $\zeta_c > 0$, that rotate the local director. This rotation is equivalent to a translation (layer displacement) along the pitch axis that opposes the initial displacement for $q_0 > 0$ (right-handed) but amplifies it for $q_0 < 0$ (left-handed).}
    \label{fig:chiral_instability}
\end{figure}

We can give another perspective on the hydrodynamics and chiral rotation-translation coupling if we restrict to displacement fields $u = u(x,y)$ that only vary transversely. In this case, and using the vector identity $\nabla^2 {\bf A} = \nabla ( \nabla \cdot {\bf A} ) - \nabla \times ( \nabla \times {\bf A} )$, the Stokes equation reduces to 
\begin{equation}
    \begin{split}
        0 & = - \nabla p - \eta \nabla \times \bigl( \nabla \times {\bf v} \bigr) + \frac{K}{2} \nabla \times \bigl( \nabla \times \nabla^2 u \,{\bf e}_z \bigr) \\
        & \quad + \frac{K}{4q_0} \nabla \times \bigl( \nabla^4 u \,{\bf e}_z \bigr) + \frac{\zeta}{2} \nabla \times \bigl( \nabla \times u \,{\bf e}_z \bigr) \\
        & \quad - \frac{\zeta_c}{2} \nabla \times \bigl( \nabla^2 u \,{\bf e}_z \bigr) .
    \end{split}
\end{equation}
The pressure is constant and we recover an expression for the vorticity 
\begin{equation}
    \begin{split}
        \nabla \times {\bf v} & = \nabla \times \biggl( \frac{K}{2\eta} \nabla^2 u + \frac{\zeta}{2\eta} \,u \biggr) {\bf e}_z \\
        & \quad + \biggl( \frac{K}{4q_0\eta} \nabla^4 u - \frac{\zeta_c}{2\eta} \nabla^2 u \biggr) {\bf e}_z ,
    \end{split}
\end{equation}
in which the first term is horizontal and the last term is the component about the layer normal. Vorticity about the layer normal corresponds to vortical flows within the pseudolayers and is a hallmark of odd elastic effects~\cite{kole2021}, see Fig.~\ref{fig:chiral_instability}. It has contributions from both the equilibrium cholesteric hydrodynamics (associated to the screw symmetry) and from the chiral activity. The direction of the former depends on the handedness of the cholesteric and is always such that the rotation is equivalent to a translation that opposes the pseudolayer displacement and acts to restore equilibrium. In contrast, the active contribution is destabilising when it has the opposite sense to the passive term and is equivalent to a translation that enhances the pseudolayer displacement, illustrated in Fig.~\ref{fig:chiral_instability}. 

From the horizontal component of vorticity (neglecting any constant part of $u$) we can also read off the normal component of velocity and then obtain the displacement field dynamics 
\begin{equation}
    \begin{split}
        \partial_t u & = - \frac{K}{2 q_0^2 \gamma} \nabla^4 u + \frac{K}{2\eta} \nabla^2 u + \frac{\zeta}{2\eta} \,u \\
        & \quad - \frac{1}{2q_0} \biggl( \frac{K}{4q_0\eta} \nabla^4 u - \frac{\zeta_c}{2\eta} \nabla^2 u \biggr) .
    \end{split}
\end{equation}
This is the real space version of~\eqref{eq:gq_approx} with $q_z = 0$. The equilibrium rotation-translation coupling term is stabilising independent of the handedness of the cholesteric, however, the active contribution ($\zeta_c$) is destabilising when the vorticity it induces is opposite in sign to the passive contribution. The criterion for instability from chiral activity ($K + \zeta_c / 2q_0 < 0$) arises from the competition between the restoring effects of the passive contribution to $v_z$ and the active contribution to $(\nabla \times {\bf v})_z$ with the rotation-translation coupling of the cholesteric screw symmetry. 

\section{Three-Dimensional Isotropic Rotation-Translation Coupling} 
\label{sec:3d_chiral_fluid}

The screw symmetry of cholesterics leads to a hydrodynamic coupling of the component of vorticity along the pseudolayer normal with its displacement field, along with the corresponding chiral force density. It is natural to contemplate that such coupling may arise in other chiral systems with screw symmetry, for example in the blue phases~\cite{wright1989,crooker1989}, which can be described by exactly the same Beris-Edwards equations for the Q-tensor order parameter as govern the hydrodynamics of the cholesteric helix. As such we speculate that the screw-symmetry chiral hydrodynamics we have described here for cholesterics will also be manifest in blue phase hydrodynamics. Although we defer a derivation of the hydrodynamics for any specific blue phase structure to subsequent work, we speculate on the isotropic (part of the) chiral hydrodynamics of such three-dimensional textures. There are three thermodynamically distinct blue phases in the absence of applied fields, two with cubic structures and one, the blue fog (blue phase III), that is amorphous~\cite{henrich2011}. While the hydrodynamics of the crystalline blue phases should reflect their cubic symmetry, the blue fog should be described by an isotropic hydrodynamics. We limit our speculations to a brief analysis of a direct generalisation of the linearised cholesteric equations to three-dimensional isotropic couplings and elasticity, which provides insight into what is possible more generally, particularly for the blue fog, and on what aspects of the cholesteric phenomenology are particular to that phase. 

As linearised hydrodynamic equations generalising those for the cholesteric, we take 
\begin{gather}
    \begin{split}
        \partial_t {\bf u} & = {\bf v} - \frac{1}{2q_0} \nabla \times {\bf v} \\
        & \quad + \frac{1}{\gamma q_0^2} \Bigl[ \bigl( \lambda + \mu \bigr) \nabla \bigl( \nabla \cdot {\bf u} \bigr) + \mu \nabla^2 {\bf u} \Bigr] , 
    \end{split} \\
    \begin{split}
        0 & = - \nabla p + \eta \nabla^2 {\bf v} + \bigl( \lambda + \mu \bigr) \nabla \bigl( \nabla \cdot {\bf u} \bigr) \\
        & \quad + \bigl( \mu - \zeta \bigr) \nabla^2 {\bf u} - \biggl( \frac{\mu}{2q_0} + \zeta_c \biggr) \nabla \times \nabla^2 {\bf u} ,
    \end{split}
\end{gather}
where $\lambda$, $\mu$ are Lam\'e coefficients for an isotropic elasticity of the (pseudo-)displacement field and $\zeta$, $\zeta_c$ are coefficients of isotropic achiral and chiral active terms directly analogous to those arising for cholesterics. The force density $\nabla\times\nabla^2{\bf u}$ is the elastic analogue of the velocity-dependent chiral force density $\propto\nabla^2\nabla\times{\bf v}$ discussed by~\cite{Andreev} for equilibrium chiral fluids. 

The structure of the linearised modes can be obtained by Fourier transform. We decompose ${\bf u}$ into components parallel, $u_{\parallel}$, and perpendicular, ${\bf u}_{\perp}$, to the wavevector ${\bf q}$ of the Fourier mode and find the dynamics 
\begin{gather}
    \partial_t u_{\parallel} = - \frac{(\lambda + 2\mu) q^2}{\gamma q_0^2} \,u_{\parallel} , \\
    \begin{split}
        \partial_t {\bf u}_{\perp} & = \frac{1}{\eta} \biggl[ \zeta - \mu - \biggl( 1 + \frac{4\eta}{\gamma} \biggr) \frac{\mu q^2}{4q_0^2} - \frac{\zeta_c q^2}{2q_0} \biggr] {\bf u}_{\perp} \\
        & \quad + \frac{i}{\eta} \biggl( \frac{2\mu - \zeta}{2q_0} + \zeta_c \biggr) {\bf q} \times {\bf u}_{\perp} .
    \end{split}
\end{gather}
The parallel mode is purely diffusive and has no contributions from either the activity or the rotation-translation coupling. Indeed, it {\sl cannot} because the velocity field is incompressible and this mode is therefore purely controlled by permeation. The orthogonal eigenmodes are helical with growth rate (${\bf u}_{\perp} \sim \mathrm{exp}(g_{\bf q}t)$) 
\begin{equation}
    \begin{split}
        g_{\bf q} & = - \frac{\mu}{\eta} \biggl( 1 - \frac{\zeta}{\mu} \biggr) \pm \frac{\mu}{\eta q_0} \biggl( 1 - \frac{\zeta}{2\mu} + \frac{\zeta_c q_0}{\mu} \biggr) q \\
        & \quad - \frac{\mu}{4\eta q_0^2} \biggl( 1 + \frac{4\eta}{\gamma} + \frac{2\zeta_c q_0}{\mu} \biggr) q^2 .
    \end{split}
\end{equation}
A notable change from the cholesteric case~\eqref{eq:gq_full} is the presence of a contribution linear in $q=|{\bf q}|$. Such a term is allowed precisely because of the rotation-translation coupling from screw symmetry. It is absent in the cholesteric because, although both the full velocity and vorticity contain terms linear in $q$, only the components along the sole displacement field of the cholesteric ({\sl i.e.} $z$) enter the dynamics and projection onto these components eliminates the linear terms. The growth rate is negative definite for large $q$ only if $2\zeta_c q_0 / \mu > - ( 1 + 4\eta / \gamma )$. Assuming this, the condition for linear instability and wavevector for the most unstable mode are  
\begin{gather}
    \biggl( 1 - \frac{\zeta}{2\mu} + \frac{\zeta_c q_0}{\mu} \biggr)^2 > \biggl( 1 + \frac{4\eta}{\gamma} + \frac{2\zeta_c q_0}{\mu} \biggr) \biggl( 1 - \frac{\zeta}{\mu} \biggr) , \\
    q^{\star} = 2q_0 \frac{\bigl| 1 - \frac{\zeta}{2\mu} + \frac{\zeta_c q_0}{\mu} \bigr|}{1 + \frac{4\eta}{\gamma} + \frac{2\zeta_c q_0}{\mu}} .
\end{gather}
As in the cholesteric case, the natural scale is that of the microstructure so that the instability is only consistently hydrodynamic close to the threshold.

\section{Discussion}
\label{sec:discussion}

In three-dimensional, chiral, orientationally-ordered materials, screw symmetry couples translational and rotational motions. At the hydrodynamic level, the rotation due to fluid vorticity about the screw axis manifests as an advection of the ordered structure along it, accompanied by an Onsager counterpart antisymmetric stress of generalised odd elastic form~\cite{lubensky1972,kole2021}. In cholesterics, this gives rise to an example of a chiral mechanical force deriving from a passive free energy, for which we present a derivation from the Ericksen-Leslie equations. Its effects are manifest only in the dynamic response and static cholesteric mechanics, determined solely from the vanishing of the hydrodynamic molecular field, remain insensitive to the handedness of the helix. The symmetry origin of these properties makes it probable that similar chiral mechanics arises in other passive materials with screw symmetry, for instance in helimagnets~\cite{radzihovsky2011} or the blue phases~\cite{wright1989,crooker1989}. Although we speculated on the isotropic form this may take, it remains to determine it explicitly, including the dependence on the particular space group. 

In active cholesterics, the screw symmetry rotation-translation coupling provides a mechanism for the chiral activity to enter the linearised pseudolayer displacement dynamics. This allows for a new mode of instability, whose signature in the growth rate is the change in sign of the term quadratic in wavevector, i.e. it is type II in the Cross-Hohenberg classification~\cite{HohenCross}. Significantly, it is, to our knowledge, the first example of an active instability that is sensitive to the handedness of the helical structure as well as that of the chiral activity. For the three-dimensional version, the analogous instability transmutes to terms linear in wavevector that are allowed precisely because of screw symmetry. Current experimental realisations of three-dimensional active nematics verify the spontaneous flow transition predicted in those systems~\cite{alam2024}. These experiments make use of {\em fd} viruses to provide a liquid crystal background for the active component and this is known to form cholesteric phases in passive systems~\cite{dogic2000}. Extension of the experimental system to the cholesteric regime should provide experimental validation of the active instabilities we have described. A separate experimental realisation is provided by swimming microorganisms in biocompatible liquid crystals such as nanocellulose~\cite{chu2022}; these experiments show the `Helfrich-Hurault' instability of the apolar active stress~\cite{whitfield2017} and should also be able to demonstrate the chiral active instabilities we focused on here.

There are numerous natural directions for further work. For example, properly establishing the hydrodynamics of materials with multiple screw symmetries, such as the blue phases, and the dependence on the point group. In addition to being relevant to active versions of these liquid crystalline phases, this may also serve to identify chiral mechanical properties of three-dimensional artificial robotic assemblages~\cite{veenstra2025}. Another interesting direction is the impact of screw symmetry on the dynamics of topological defects, such as $\lambda$ and $\chi$ lines in cholesterics, or of other chiral topological solitons~\cite{pollard2025}. The analysis we have given here applies outside a core region about each defect and it would be interesting to determine how it matches onto a separate treatment of the defect core. Chirality is ubiquitous and it seems likely that chiral hydrodynamical phenomena arise naturally in a diverse range of biological settings. 

\begin{acknowledgments}{
AM was supported in part by a TALENT Fellowship awarded by CY Cergy Paris Universit\'e and an ANR grant, PSAM. SR was supported in part by a J C Bose Fellowship of the ANRF, India, and an Endowed Visiting Professorship of ICTS-TIFR. SJK was supported by the INI-Simons Research Fellowship, Cambridge. This research was supported in part by Grant No.~NSF PHY-2309135 to the Kavli Institute for Theoretical Physics.}
\end{acknowledgments}

\bibliographystyle{unsrt}
\bibliography{ref}

\end{document}